\documentclass[namedreferences]{SolarPhysics}
\usepackage{setspace}
\usepackage[optionalrh]{spr-sola-addons} 
\usepackage{graphicx}                    
\usepackage{color}                       
\usepackage{url}                         

\newcommand{\etal}{{\it et al.}}

\newcommand{\aap}{    {\it Astron. Astrophys.}}

\newcommand{\apj}{    {\it Astrophys. J.}}
\newcommand{\apjl}{    {\it Astrophys. J. Lett.}}

\newcommand{\mnras}{  {\it Mon. Not. Roy. Astron. Soc.}}
\newcommand{\nat}{    {\it Nature}}

\newcommand{\solphys}{{\it Solar Phys.}}

\begin{document}

\begin{article}

\begin{opening}

\title{Comparisons of Supergranule Characteristics During the Solar Minima of Cycles 22/23 and 23/24}

%
\author{Peter~E.~\surname{Williams}$^{1}$\sep
        W.~Dean~\surname{Pesnell}$^{2}$
       }

%
\runningauthor{P.~E.~Williams~and~W.~D.~Pesnell}
\runningtitle{Comparisons of Supergranules During Solar Minima 22/23 and 23/24}

%
  \institute{$^{1}$ NASA Postdoctoral Program Fellow,\\
  Code 671, NASA Goddard Space Flight Center, Greenbelt, MD, USA
\\ email: \url{peter.williams@nasa.gov} \\ 
             $^{2}$ Code 671, NASA Goddard Space Flight Center, Greenbelt, MD, USA
\\email: \url{william.d.pesnell@nasa.gov}\\
             }

Supergranulation is a component of solar convection that manifests itself on the photosphere as a cellular network of around 35 Mm across, with a turnover lifetime of 1--2 days. It is strongly linked to the structure of the magnetic field. The horizontal, divergent flows within supergranule cells carry local field lines to the cell boundaries, while the rotational properties of supergranule upflows may contribute to the restoration of the poloidal field as part of the dynamo mechanism that controls the solar cycle. The solar minimum at the transition from cycle 23 to 24 was notable for its low level of activity and its extended length. It is of interest to study whether the convective phenomena that influences the solar magnetic field during this time differed in character to periods of previous minima. This study investigates three characteristics (velocity components, sizes and lifetimes) of solar supergranulation. Comparisons of these characteristics are made between the minima of cycles 22/23 and 23/24 using MDI Doppler data  from 1996 and 2008, respectively. It is found that whereas the lifetimes are equal during both epochs (around 18 h), the sizes are larger in 1996 (35.9 $\pm$ 0.3 Mm) than in 2008 (35.0 $\pm$ 0.3 Mm), while the dominant horizontal velocity flows are weaker (139 $\pm$ 1 m s$^{-1}$ in 1996; 141 $\pm$ 1 m s$^{-1}$ in 2008). Although numerical differences are seen, they are not conclusive proof of the most recent minimum being inherently unusual.

%

\end{opening}

%

 \section{Introduction} \label{S:Intro}

Studies of the solar cycle are important for understanding
hydrodynamic mechanisms within the Sun. They are increasingly being
used for making predictions in the long-term about upcoming cycles
and in the short-term about evolution of active regions. Such predictions have
 become important with advances in human technology
that can be affected by space weather for which the Sun is ultimately
responsible. Studies throughout and comparisons between
cycles are vital in constraining these predictions. With the
{\it Solar and Heliospheric Observatory} (SOHO) mission, the
opportunity has arisen, for the first time, to make solar
observations from a single, space-based instrument through consecutive solar
minima. This lays the path for making direct comparisons between
adjacent cycles that will assist in prediction making.

Convection is inextricably linked to processes that determine the
behavior of the solar cycle via its interaction with the magnetic
field. Internal convective motions induce the lifting and twisting
of field lines (the $\alpha$-effect) that is at the core of many
dynamo models (for example, \opencite{Kapyla06}). It is therefore worthwhile
to study convective processes to understand their
influence on the local magnetic field that may also play a cumulative
role in structuring the global field. On the other hand, it is of
further interest to understand how convection processes change due
to the varying strength of the global field over the course of a
solar cycle.

Convection has a variety of components, two well-studied examples being
granulation and supergranulation, although the convective nature of the latter
is not conclusive. For example, \inlinecite{Rieutord00} have suggested that supergranular flow is
generated directly by the granular flow, \inlinecite{Beck00}, among others, suggest a wave propagation
element to supergranulation, while \inlinecite{Rast03} has proposed that supergranule scales
arise through the collective interaction of many small-scale and short-lived granular downflow plumes.
Granulation is well resolved in white light observations as bright cellular structures surrounded by
dark lanes. They measure about 1 Mm across and have a typical lifetime of around
an hour. Their larger counterparts, supergranulation, are typically around
35 Mm across and last for anywhere between 24--48 h. Initially
discovered by \inlinecite{Hart56} as velocity fluctuations over the mean rotation rate,
they were first measured using Doppler imaging methods by
\inlinecite{Leighton62} which remains a reliable means of observing
supergranulation to the present day. Mesogranulation ($\approx$ 7 Mm 
across) and giant cells ($\approx$ 100 Mm across) have been
proposed to exist (\opencite{November81}; \opencite{Simon68}), 
but no hard evidence has been forthcoming. It is
possible that mesogranulation exists as a combination of small
supergranules and large granules, as opposed to a distinct component
of convection in itself. Indeed, \inlinecite{Hathaway00} and \inlinecite{Rieutord08} see no appearance
of a mesogranulation component in the power spectra and \inlinecite{Matloch09} suggests that they are a 
result of the averaging of random data. The direct existence of giant cells has
also proved difficult to establish although indirect correlation
methods (for example, \opencite{Beck98}) have discovered latitudinally-elongated
structures that exist over many solar rotations. Giant cells have also
been studied via numerical simulations, which may assist in their
understanding and direct observation (\opencite{Miesch00}; \opencite{Williams09}).

As material from below the photosphere is convected
radially upward by the supergranule cells, magnetic field lines are
carried along with it. On reaching the surface, the supergranule
flow becomes horizontally dominated. This carries the field lines to
the edges of the supergranule cell where the field lines are trapped at stagnant points of convergent flow and
are forced to advect around the cell boundaries. Supergranulation is
observed to be directly linked to the magnetic field via field
elements that trace out the boundaries of supergranules clearly
seen within, for example, photospheric magnetograms and Ca {\sc II} K
observations of the chromospheric network. H$\alpha$
observations also show these cellular structures, as well as evidence of
chromospheric filament material accumulating at the boundaries of
supergranules \cite{Pevtsov05}.

The Michelson Doppler Interferometer (MDI) instrument aboard SOHO \linebreak
\cite{Scherrer95} has produced approximately 60-days
of full-disk Dopplergrams each year since 1996, thus covering a
full solar cycle. Analysis of MDI data has produced a vast array of information about
supergranulation such as their evolution \cite{DeRosa04}, alignment
\cite{Lisle04} and rotation around the Sun \cite{Meunier07}  and
aided in constraining both numerical simulations of convection \cite{Stein00} and data simulations
that produce realistic Dopplergrams by modeling the convection spectrum of MDI \cite{Hathaway06}.

The most recent solar minimum ({\it i.e.}, the transition from cycle 23 to 24) has
provided much interest due to its extended nature. This has been quantified by 
characteristics such as the sunspot number which remained at zero over a period of
time unprecedented in the current epoch. Many comparisons between this and previous
minima are being made (for example, \opencite{Howe09}) to determine whether this 
minimum stands out as being peculiar.

This work contributes to the comparison between the solar minimum during the transition from cycle 22 to 23, with that from cycle 23 to 24. It uses SOHO/MDI data that covers both periods of interest. It will add to the ongoing study of any peculiarities, as well as the causes and consequences of the 23/24 minimum. By studying supergranule characteristics during both epochs, changes in the convection zone that reflect the differences in the behavior of the magnetic field at these two times may be quantified.

\section{Data Preparation} \label{S:DataPrep}

A popular method of studying supergranule convection flows is
through the use of Dopplergrams. In recent years, these have been predominantly
delivered by MDI upon SOHO. MDI has produced yearly
Dynamics Runs of approximately 60-days worth of 1024$\times$1024
pixel Dopplergrams at a 1-min cadence. For this study, 62 days of data from
1996 (MDI Day 1239-1300; 24 May 24 1996 - 24 July 1996) and 60 days from 2008 (MDI Day 5540-5599; 3 March 2008 - 1 May 2008), 
both during a period of solar minimum, are used.

To extract convective kinematic characteristics, the Dopplergrams have been
reduced to isolate the line-of-sight velocities of surface convection
cells. From these, a variety of analysis techniques are used to
determine values for the horizontal and radial velocity components
of supergranules, their spatial sizes and their 1/e lifetimes,
and to provide inter-year comparisons.

A number of independent velocity signals need to be removed from
each Dopplergram to extract supergranule flow velocities.
First to be removed are the velocities arising due to 5-min 
p-mode oscillations \cite{Hathaway88}. A 31-min temporal Gaussian filter, with an
FWHM of 8 min, is constructed to produce a weighted average of a set of 
31 1-min Dopplergrams. Each component Dopplergram is de-rotated
to match the convection pattern with that of the central Dopplergram within the filter. The resultant
time-series is sampled every 15 min, producing a set of 96
averaged Dopplergrams per day.

After subtracting the gravitational redshift and velocity
signals due to the motion of the observer, each averaged Dopplergram
undergoes further reduction with the removal of the axisymmetric
flows (\opencite{Hathaway87}, \citeyear{Hathaway92}). Fits are
produced for the velocity fields relating to the differential
rotation, meridional circulation and convective blueshift due to unresolved convection components, 
that are subsequently subtracted from the Dopplergrams. The resulting velocity images contain only the line-of-sight
velocity fields at the photosphere which may be displayed either in
heliocentric or heliographic coordinate systems depending on the analysis.

During the removal of the aforementioned velocity fields, each
Dopplergram is mapped to a heliographic coordinate system and the resulting
images are projected onto the spherical harmonics, following the
method of \inlinecite{Hathaway00}. This produces a two-dimensional power
spectrum in spherical harmonic degree, $\ell$, and order, $m$. 
If necessary, contributions from instrumental calibration artifacts
may be removed by setting the power at $\ell$ $<$ 11, for all $m$, equal to zero.

  \begin{figure}        
     \centerline{\includegraphics[width=0.9\textwidth]{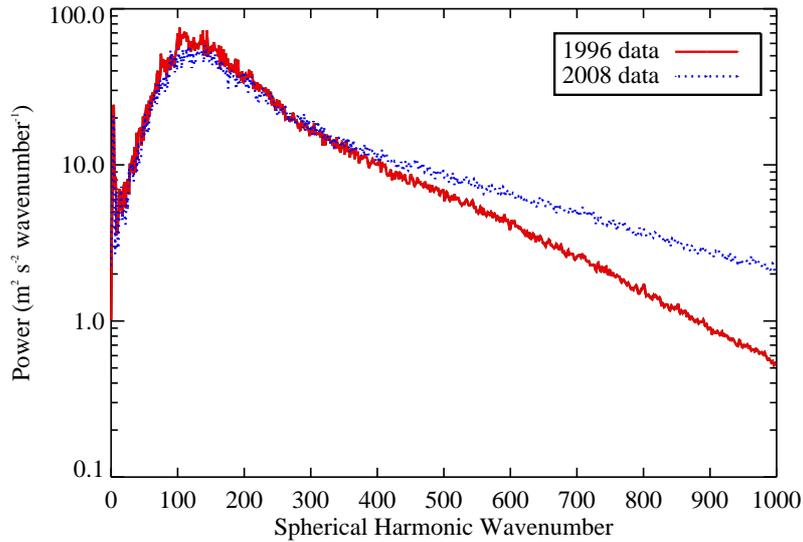}}
\caption{Examples of one-dimensional power spectra derived from
processed Dopplergrams (solid line - 0000 UT, 25 May 1996;  dotted line - 0000 UT, 4 March 2008). The supergranule feature that peaks at around $\ell$ = 125 dominates the spectrum. Power at low wavenumbers is related to instrument calibration artifacts. The reduction in power at high wavenumbers is due to the relative lack of features at that spatial scale, compared to supergranulation. Note that the reduction in power is steeper for 1996 than for 2008. This is caused by changes in defocusing between 1996 and 2008 \protect\cite{Korzennik04} that impacts the wavenumber domain above $\ell$ $\approx$ 300.}
   \label{Fig:SpecEx}
   \end{figure}

Additionally, the two-dimensional power spectrum is
summed over $m$ to produce a power spectrum in one-dimension
(Figure~\ref{Fig:SpecEx}). The strong feature seen peaking at
around $\ell$ = 125 is caused by the presence of supergranules
with a typical diameter of around 35 Mm.

\section{Data Analyses} \label{S:Analyses}
\subsection{Supergranule Sizes} \label{ss:SgSize}

Supergranule sizes have been studied via analyses of magnetograms \cite{Komm95}, Ca {\sc II} K images \cite{Hagenaar97} and intensity maps \cite{Meunier08}. The method adopted in this paper to extract global averages of supergranule sizes from full-disk Dopplergrams is described by \inlinecite{Hathaway00} and examples of the spectra produced are shown in Figure \ref{Fig:SpecEx}.

Fits to the supergranule component of the convection
spectrum can parameterize supergranule sizes. This process derives
the wavenumber at which the peak occurs, which will provide a typical
supergranule size estimation. Furthermore, a range of typical sizes can be
determined by calculating the FWHM of the supergranule spectral peak.

\begin{figure}        
     \centerline{\includegraphics[width=0.9\textwidth]{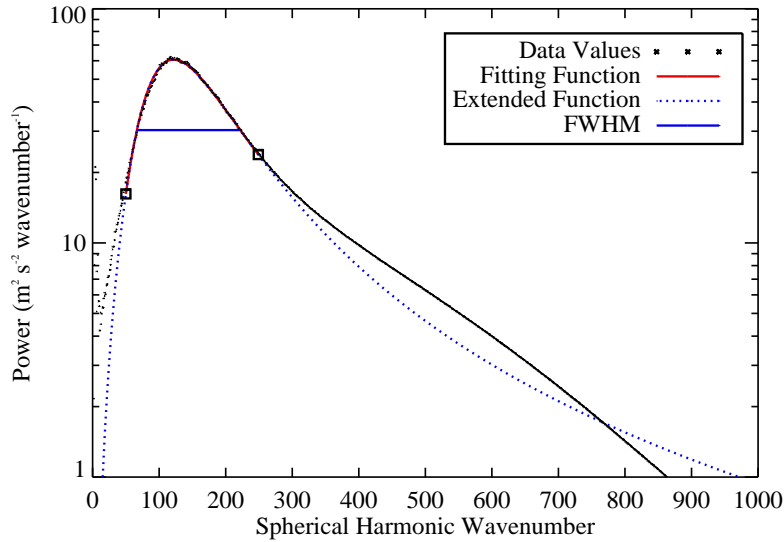}
              }
\caption{An averaged convection spectrum is fitted by a modified Lorentzian
between 50 $\leq \ell \leq$ 250. The fit, performed within the range represented by the square symbols,
provides both the wavenumber at which the peak
occurs and the FWHM of the feature. For the 1996 dataset shown, these were found to be 
$\ell_{\rm{peak}}$ = 122 $\pm$ 1 and $\Delta\ell$ = 155 $\pm$ 3, respectively. For the 2008 dataset,
$\ell_{\rm{peak}}$ = 125 $\pm$ 1 and $\Delta\ell$ = 164 $\pm$ 5.
        }
   \label{Fig:Sizes}
   \end{figure}

We have analyzed the spectra from 1996 and 2008. For each year,
spectra were produced for every Doppler image and averaged over the dataset to
produce a spectrum with reduced noise and a smooth distribution. We
performed fits to the supergranule peak, between the wavenumber
range 50 $\leq \ell \leq$ 250 of the convection spectrum, using a
variety of fitting functions. By analyzing the residuals between the
fit and the data between the given wavenumber range, we found that a
modified Lorentzian, given by
 \begin{equation}  \label{Lorentzian}
    f(\ell) = \left\{\frac{\ell A}{[(\ell-\ell_{0})^{2} + \Gamma^{2}]}\right\}^{2},
 \end{equation}

\noindent provides the best fit to the peak (Figure~\ref{Fig:Sizes}), as opposed to the
Boltzmann and log-normal functions that were also tested, where $\ell$ is the wavenumber,
and $A$, $\ell_{0}$, and $\Gamma$ are the amplitude, peak wavenumber and width
of the Lorentzian, respectively. A typical supergranule diameter, $\lambda$, may be derived from the peak wavenumber
of the fit, $\ell_{\rm{\rm{peak}}}$, using
 \begin{equation}  \label{SGsizes}
    \lambda = \frac{2\pi R_{\odot}}{\ell_{\rm{\rm{peak}}}},
 \end{equation}
which may also be used to determine a spatial range, $\Delta\lambda$, from the end values of the FWHM.

For the 1996 dataset, the peak wavenumber was determined to be
$\ell_{\rm{peak}}$ = 122 $\pm$ 1, with a FWHM of $\Delta\ell$= 155 $\pm$ 3,
corresponding to an average diameter of $\lambda$ = 35.9 $\pm$ 0.3 Mm and a size range of
19.7--65.6 Mm ($\Delta\lambda$ = 45.9 Mm). The 2008 dataset gave a peak wavenumber of $\ell_{\rm{peak}}$ =
125 $\pm$ 1 and a FWHM of $\Delta\ell$= 164 $\pm$ 5, corresponding
to an average diameter of $\lambda$ = 35.0 $\pm$ 0.3 Mm and a size range of 18.9--65.0 Mm ($\Delta\lambda$ = 46.1 Mm).

\subsection{Supergranule Velocity Components} \label{ss:SgVelComp}

Below the surface, convecting material may be assumed to be rising
vertically, carrying energy outward from the solar interior. Near the surface, where the 
material becomes convectively stable, the radial flow becomes a
divergent, horizontal flow and the material spreads out over the surface from the
center of the convection cell with the excess energy radiated to space. 
Where adjacent cells meet, their respective horizontal flows converge at their 
boundaries, becoming vertical downflows.

It is of interest to measure these velocity components as both flows
play an important role in structuring the magnetic field. Radial flows tend to bring
field lines to the surface, while horizontal flows advect the lines to the edges of the
surface cells to produce structures such as those seen in the chromospheric network.

\inlinecite{Hathaway02} devised a method to extract the RMS velocity
components by analyzing Dopplergrams from the 1996 MDI Dynamics Run.
They derived typical values for radial and horizontal flow
components of $V_r$ = 29 $\pm$ 2 m s$^{-1}$ and $V_h$ = 258 $\pm$ 1
m s$^{-1}$, respectively. We have applied this analysis to both the 1996 and 2008 MDI Dynamics
Run data to compare typical values of these velocity components. The
method is as follows.

Each Dopplergram is reduced to a heliocentric map of the supergranulation velocity
as described in Section~\ref{S:DataPrep}. Each pixel's velocity value is squared and the resulting array
is binned into a set of 200 annuli centered on disk center at a heliocentric angular distance,
$\rho$, from disk center. The average squared velocity in each annulus is then calculated.
This process is repeated for each image in the data set and the average over the dataset for each annulus is taken.
This produces a set of mean-squared velocity values, $\overline{V^{2}(\rho)}$ at angular distance, $\rho$.

Using \cite{Hathaway02},
 \begin{equation}  \label{LinearEqn}
    \overline{V^{2}(\rho)} = \overline{V^{2}_{r}} + (\overline{V^{2}_{h}}-\overline{V^{2}_{r}})\sin^{2}\rho,
 \end{equation}
 
\noindent where $V_r$ and $V_h$ are the radial and horizontal
velocity components, respectively, the mean-squared velocity values can
then plotted against $\sin^{2}\rho$. It should be noted that, as well as $V_r$, the $V_h$ component is
limited to velocities directed along the line of site; any
horizontal component perpendicular to the line of sight does not
contribute. Linearly fitting the data provides $V_r$ and $V_h$ from
the intercept and gradient of the fit, respectively. The fit is
applied out to $\sin^{2}\rho$=0.25 ($\rho$ = 30$^{\circ}$) as, above this value, the data falls away
due to limb effects. These methods were applied to both the 1996 and 2008 
datasets and the results compared.

For 1996, the results derived were $V_r$ = 24 m s$^{-1}$ and $V_h$ =
278 m s$^{-1}$, with a $V_r$/$V_h$ ratio of 0.09, values similar to
\inlinecite{Hathaway02}. For 2008, the analysis produced $V_r$ = 40
m s$^{-1}$ and $V_h$ = 303 m s$^{-1}$ with a $V_r$/$V_h$ ratio of
0.13, values notably greater than for 1996. However, this discrepancy
is explained by the large amount of optical defocusing of MDI during 1996
\cite{Korzennik04} which tends to blur the Doppler image and
reduce the velocity value contained in each image pixel. The effect
is also seen in the convection spectrum where the high wavenumber
tail is lower in power for 1996 than for 2008 (Figure \ref{Fig:SpecEx}). We have attempted to
make the two datasets equivalent by; (1) smearing the
2008 Doppler images to reduce the pixel velocities and thus the
spectral power at high wavenumbers; and (2) spectrally filtering the
images of both 1996 and 2008 so that power at the high wavenumber
tail does not contribute to the velocity analysis so removing the influence
of defocusing.

Dopplergrams from 2008 underwent smearing to mimic the defocusing of
the 1996 images. This was carried out on the original 15-min
cadence images prior to the reduction process so that the artificial
defocusing could be performed as near to the source of the images as
possible. The smeared image is then processed as described in
Section~\ref{S:DataPrep} and the respective convection spectrum
compared to one from 1996. A two-dimensional Gaussian with a FWHM of 3.5 pixels
smeared the images by an amount that resulted in spectra at high-wavenumbers
very similar to those from 1996. The velocity
analysis was then carried out on the smeared 2008 data with the
linear fit applied out to $\sin^{2}\rho$ = 0.36 ($\rho$ $\approx$ 37$^{\circ}$), giving
velocity components of $V_r$ = 28 $\pm$ 3  m s$^{-1}$ and $V_h$ =
269 $\pm$ 3  m s$^{-1}$ with a $V_r$/$V_h$ ratio of 0.11 $\pm$ 0.01.
For a similar fitting regime, for 1996 the values are $V_r$ = 26
$\pm$ 2 m s$^{-1}$ and $V_h$ = 272 $\pm$ 2  m s$^{-1}$ with a
$V_r$/$V_h$ ratio of 0.11 $\pm$ 0.01. However, comparing the two spectra at
high wavenumbers exhibited noticeable differences in power which would still contribute
to the velocity values, thus this method was deemed unsatisfactory for making a decent comparison
of velocities between the two datasets.

The other method to equate the data sets and remove the influence of
the defocusing is by selecting velocities within a limited range of
the convection spectrum, notably around the supergranulation peak,
thus suppressing the power at high wavenumbers (Figure~\ref{Fig:SpecEx}).
We have filtered the 1996 and 2008 reduced Doppler images with a
log-normal distribution given by

 \begin{equation}  \label{FilterEqn}
    F(\ell) = \frac{1}{\sigma\ell\sqrt{2 \pi}}\exp\left[{\frac{-\ln^{2}(\ell/\mu)}{2 \sigma^{2}}}\right],
\end{equation}
\\
\noindent where $\ell$ denotes the wavenumber, and $\mu$ and $\sigma$ are the mean and
standard deviation of the variable, respectively. These latter parameters are adjusted so that the filter
retains much of the power within the supergranule feature, but reduces power at very low wavenumbers (to remove
any remaining instrumental artifacts) and at higher wavenumbers so to remove the discrepancy between the
1996 and 2008 spectra. For this study, the values used were, $\mu$ = 136 for 1996 and $\mu$ = 139 for 2008 to
account for the shift in the mean peak between the two datasets as discussed in Section~\ref{ss:SgSize}, while
$\sigma$ = 0.35 for both years. The function was normalized by dividing through it by its maximum value.

\begin{figure}    

               \includegraphics[width=1.0\textwidth,clip=]{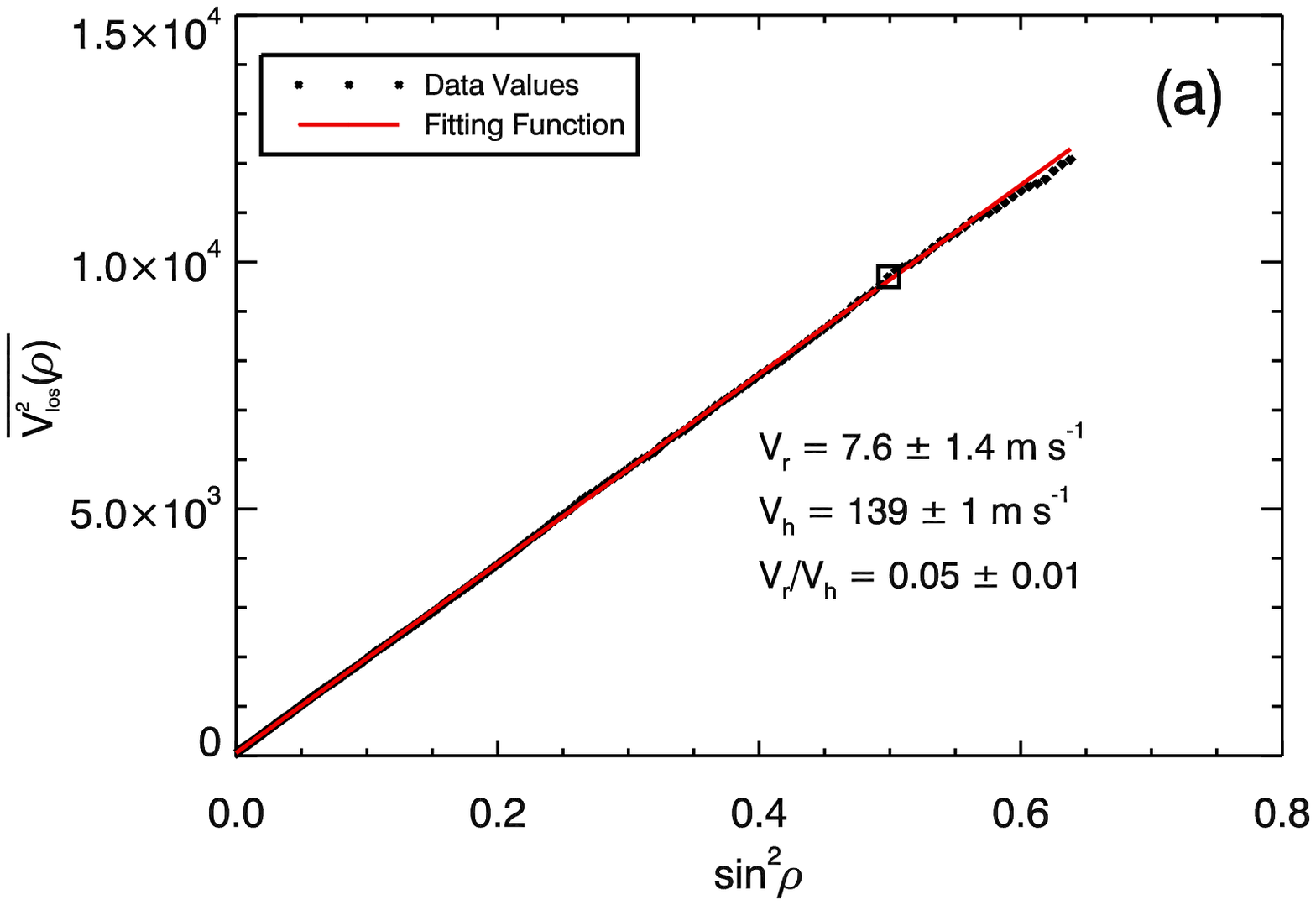}
     \vspace{0.31\textwidth}    
%
               \includegraphics[width=1.0\textwidth,clip=]{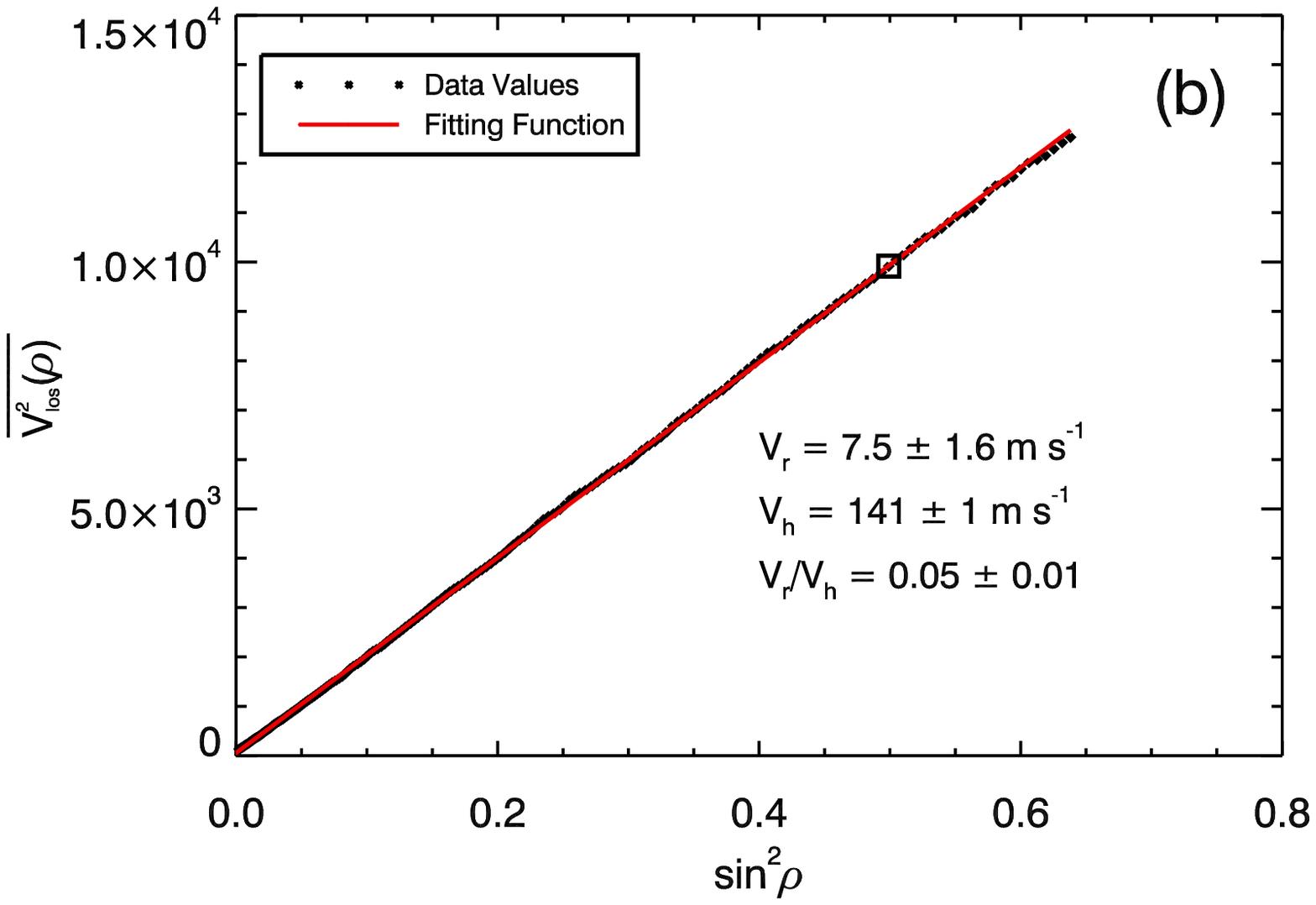}
     \vspace{-0.35\textwidth}   

\caption{Plots of mean-squared line-of-sight velocity, $\overline{V^{2}(\rho)}$, as a function of $\sin^{2}\rho$, where $\rho$ is the heliocentric angular distance from disk center, averaged over a whole dataset for 1996 (\emph{a}) and
2008 (\emph{b}). The linear fits were applied out to $\sin^{2}\rho$ = 0.5, signified by the square symbol. Above this value, limb effects begin to non-linearly affect the data. For 1996 (\emph{a}), 
we get $V_r$ = 7.6 $\pm$ 1.4  m s$^{-1}$ and $V_h$ = 139 $\pm$ 1  m s$^{-1}$ with a $V_r$/$V_h$ ratio of 0.05
$\pm$ 0.01. For 2008 (\emph{b}), we find $V_r$ = 7.5 $\pm$ 1.6  m s$^{-1}$ and
$V_h$ = 141 $\pm$ 1  m s$^{-1}$ with a $V_r$/$V_h$ ratio of 0.05 $\pm$ 0.01. As the horizontal velocity provides
most of the power, this is the best parameter with which to make inter-year comparisons.} 
   \label{VelocityFigures}
   \end{figure}

We then performed the same analysis on these filtered images, but extending the linear fit out to $\sin^{2}\rho$ = 0.5 ($\rho$ = 45$^{\circ}$) as the filtering tends to reduce the center-to-limb effects. For 1996 (Figure~\ref{VelocityFigures}\emph{a}), we get $V_r$ = 7.6 $\pm$ 1.4  m s$^{-1}$ and $V_h$ = 139 $\pm$ 1  m s$^{-1}$ with a $V_r$/$V_h$ ratio of 0.05 $\pm$ 0.01. For 2008 (Figure~\ref{VelocityFigures}\emph{b}), we find $V_r$ = 7.5 $\pm$ 1.6  m s$^{-1}$ and $V_h$ = 141 $\pm$ 1  m s$^{-1}$ with a $V_r$/$V_h$ ratio of 0.05 $\pm$ 0.01.

\subsection{Supergranule Lifetimes} \label{ss:SgLife}

Convection cells have a limited lifetime. While it is observed that granules last much less than an hour
\cite{Hirzberger99}, supergranules decay after about 24--48 h \cite{Duvall80}
 and giant cells can survive up to a few solar rotations \cite{Beck98}. 
Analyzing a time-series of Doppler velocity images,
it can be seen that the patterns change with time as they move
across the solar disk due to rotation. By following the rotation,
correlation methods can be used to track the pattern and determine
its decay rate.

\begin{figure}        
     \centerline{\includegraphics[width=1.0\textwidth,clip=,]{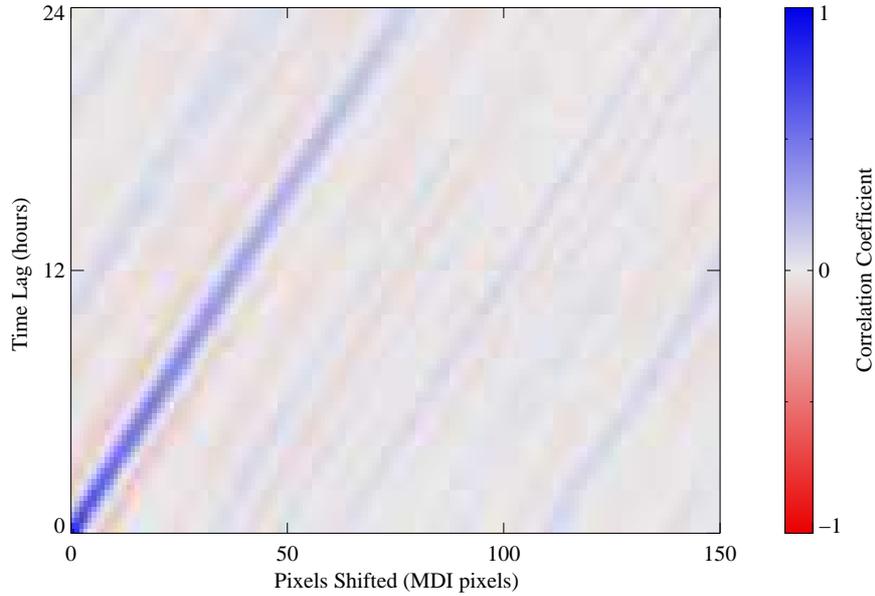}
              }
\caption{A correlation array for MDI Day 5541 (4 March 2008) dataset. The diagonal
path from the bottom left corner shows decaying positive correlation tracking the evolution
of the original supergranule Doppler pattern through subsequent images. After using Radon
transforms to produce a mask that isolates the correlation path, calculating the peak
correlation at each time-lag provides a dataset of decaying correlation values for a given day.
        }
   \label{Fig:Life2D}
   \end{figure}

Our correlation method is an adaptation of the one used by \inlinecite{Hathaway06} to study the superrotation of
supergranules. After remapping to heliographic coordinates, we select a strip of velocity data containing a supergranule Doppler velocity pattern from the time-series. On each subsequent heliographic image, another strip of
data the same size as the original strip is moved pixel by pixel
longitudinally across the image and the correlation between the
shifted strip and the original strip calculated. This is done for
every image in the time series.  The strips are 300 pixels in length by 50 pixels in height corresponding to $\approx$ 27$^{\circ}$ in longitude by $\approx$ 9$^{\circ}$ in latitude. They are extracted near the equator (centered on $\approx$ 4.5$^{\circ}$N) to reduce perspective effects and positioned in the western hemisphere so that correlations are not affected by the sign change in the velocity values as the tracked patterns cross the central meridian. The result is that for every 15-min image, a one-dimensional array of correlation coefficients for every pixel shifted is produced. The result is a two-dimensional (pixel-shift vs. time-lag) array of correlation coefficients (Figure~\ref{Fig:Life2D}).

The decrease in the correlation coefficient with respect to the
original image can be analyzed to follow the decay of a selected
convection pattern. The correlation values can be tracked in time,
beginning from the bottom left corner of the array and following a
track of correlation values. It can be seen from Figure~\ref{Fig:Life2D} that the
correlation follows a diagonal path of positive
correlation from the bottom-left of the array. This shows that the
convection pattern moves, due to rotation, across the image in time as it decays. A Radon transform can be used to select and isolate features within images by parameterizing straight lines in image space to corresponding points in Radon space defined by the gradient and intercept of the image lines.
Performing a Radon transform on the correlation array and selecting a region in
Radon space at intercept values on either side of zero at the
gradient value of the diagonal track can be used to isolate the
track. These values can be used to produce an array in Radon space
that, when inversely transformed, produces a mask that can be
overlaid over the correlation array to select the diagonal
track and mask out all other correlation values. The maximum at
each time lag in the correlation of the track can then be found. The
result is a correlation coefficient for each time-lag.

An alternative method for following the correlation track is to search for a local maximum within a limited pixel range ($\pm 10$) around the position of the maximum for the previous time-lag. This maximum would give the correlation coefficient for that time lag. Both of these methods produced identical results, as expected, and helped to isolate the correlation maximum for each time-lag, while avoiding any false maxima that might otherwise have been inadvertently selected.

It is found that the extracted correlation coefficient, $C(\Delta\tau$) decays
exponentially with lag-time, $\Delta\tau$. The slope of a straight line fitted to a log-linear
plot ($\ln C[\Delta\tau]$ vs. $\Delta\tau$) gives the decay coefficient. The inverse of this slope provides the
1/e lifetime, $\tau_{\rm 1/e}$, of the pattern.

  \begin{figure}    

               \includegraphics[width=1.0\textwidth,clip=]{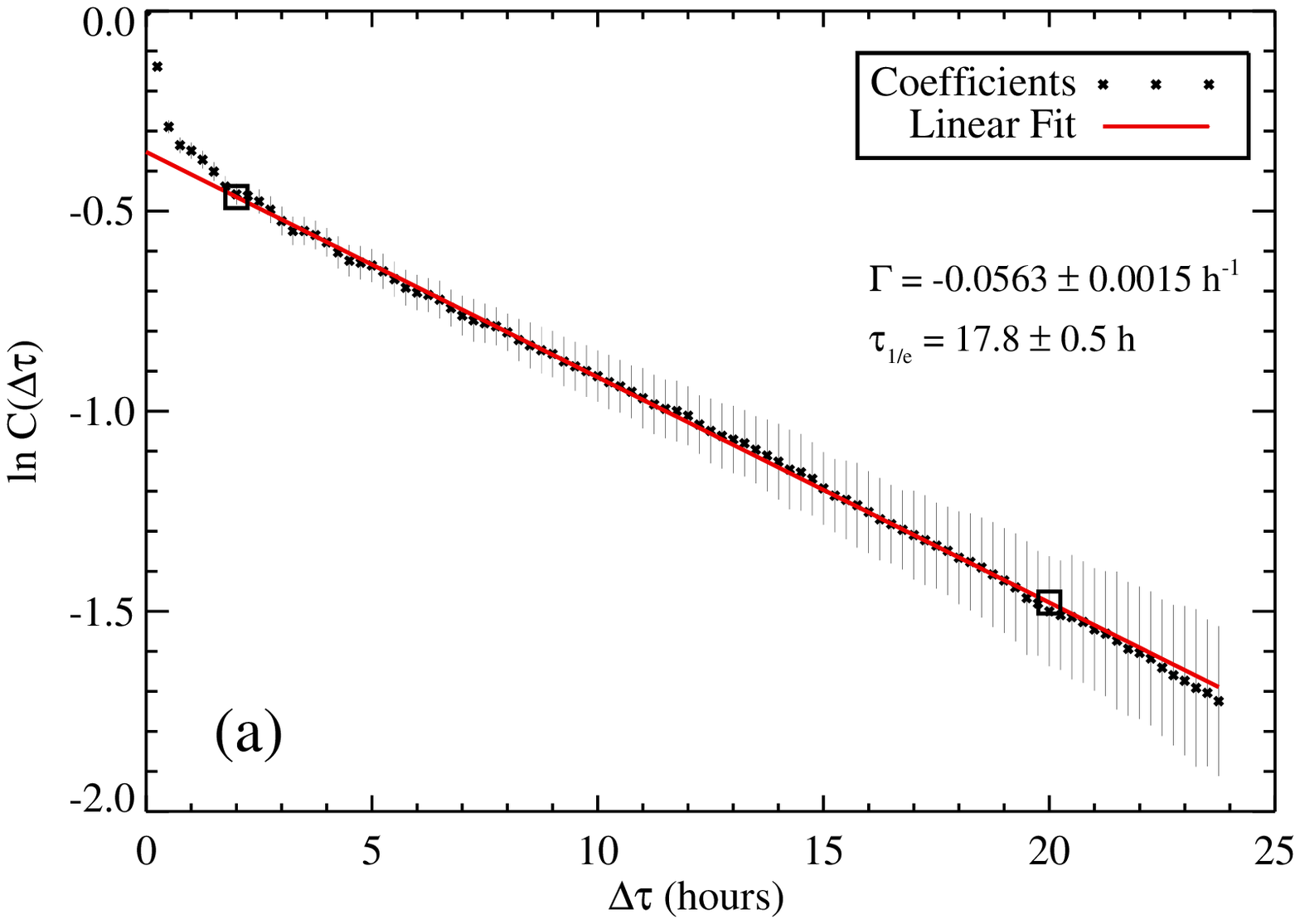}
     \vspace{0.31\textwidth}    
%
               \includegraphics[width=1.0\textwidth,clip=]{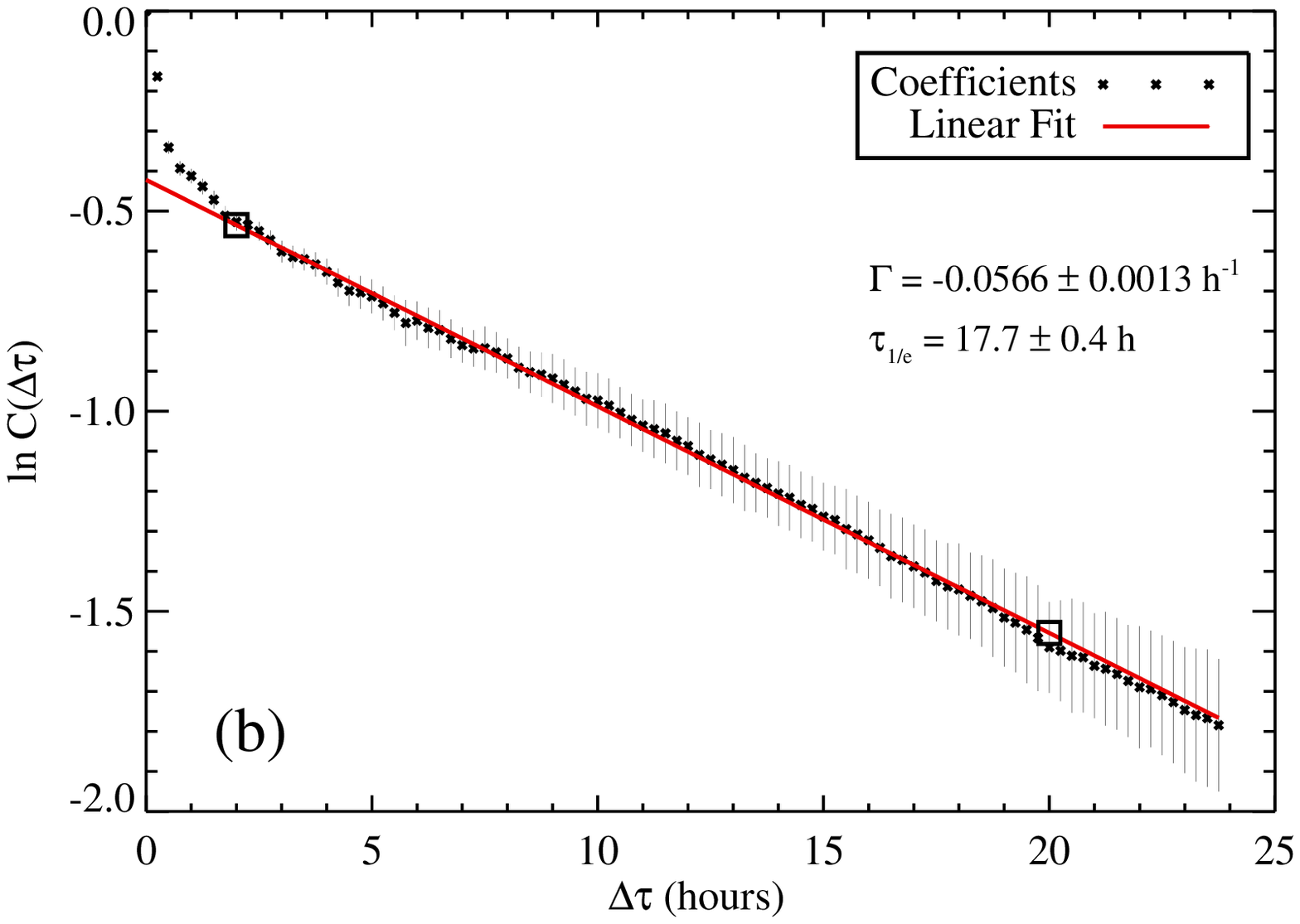}
     \vspace{-0.35\textwidth}   

\caption{Correlation coefficients, $C(\Delta\tau)$, at each time-lag, $\Delta\tau$, are
extracted from the correlation path shown in Figure~\ref{Fig:Life2D} for each
day and averaged over the whole year's dataset to produce the $C(\tau)$ vs $\tau$ data shown here; (a) and (b) show the respective 1996 and 2008 data averages and fits. The data is linearly fitted and the gradient provides the decay coefficient, $\Gamma$, from which the 1/e lifetime, $\tau_{\rm 1/e}$, is derived. In both figures, the points show the data that is fitted by the solid line between a given time range indicated by the square boxes. For both 1996 and 2008 the lifetimes are determined, from the decay coefficient, $\Gamma$, to be 18 h.
        }
   \label{Fig:Life}
   \end{figure}

This process was performed for each day within both
datasets. A strip extracted from the first image within a given
day is correlated with a similar strip moved over the subsequent set
of 95 images. Correlation coefficient vs. time-lag data is averaged
over the entire dataset and a linear fit produced. This fit provides
the the decay coefficient and 1/e lifetime for each dataset (Figure~\ref{Fig:Life}). The faster-than-exponential decay at early times in both plots is due to the presence of granules with lifetimes of 1-2 h (D.H. Hathaway, private communication).

The 1996 dataset provides an average decay coefficient, $\Gamma$ =
5.63 $\pm$ 0.15 $\times$ 10$^{-2}$ h$^{-1}$ or $\tau_{\rm 1/e}$ = 17.8 $\pm$ 0.5 h.
For 2008, the average decay coefficient, $\Gamma$ = 5.66 $\pm$
0.13 $\times$ 10$^{-2}$ h$^{-1}$ or $\tau_{\rm 1/e}$ = 17.7 $\pm$ 0.4 h.

 \section{Discussion, Conclusion and Future Work} \label{s:DiscConFut}

A summary of the parameters compared between 1996 and 2008 is given in Table~\ref{Parameters}. An additional parameter is the average number of supergranules on the solar disk during each year which is calculated using
 \begin{equation}  \label{FilterEqn}
    N_{\rm SG} = \frac{4 \pi R^{2}_{\odot}}{\pi \left(\lambda/2\right)^2} = 16 \left(\frac{R_{\odot}}{\lambda}\right)^{2},
\end{equation}
\noindent where $R_{\odot}$ is the solar radius and $\lambda$ is the average supergranule diameter. It is found that there is are 5\% increase in number of supergranules on the solar disk from 1996 to 2008.
\\
\begin{table}[htbp]
{\small
\begin{tabular}{ l l l r r }
\hline
\multicolumn{1}{c}{Parameter}&\multicolumn{1}{c}{Symbol}&\multicolumn{1}{c}{Units}&\multicolumn{1}{c}{1996}&\multicolumn{1}{c}{2008}\\
\hline
Spectral peak wavenumber & $\ell_{\rm{peak}}$ & & $122 \pm 1$ & $125 \pm 1$ \\
Spectral peak FWHM & $\Delta\ell$ & & $155 \pm 3$ & $164 \pm 4$ \\
Supergranule diameter & $\lambda$ & Mm & $35.9 \pm 0.3$ & $35.0 \pm 0.3$ \\
Radial velocity & $V_{r}$ & m s$^{-1}$ & $7.6 \pm 1.4$ & $7.5 \pm 1.6$ \\
Horizontal velocity & $V_{h}$ & m s$^{-1}$ & $139 \pm 1$ & $141 \pm 1$ \\
1/e Lifetime & $\tau_{\rm 1/e}$ & h & $17.8 \pm 0.5$ & $17.7 \pm 0.4$ \\
Number of SGs & $N_{\rm SG}$ & & $6005$ & $6318$ \\
\hline
\end{tabular}
}
\caption{Table comparing supergranule parameters from 1996 and 2008.}
\label{Parameters}
\end{table}

By averaging all the power spectra derived from all the Dopplergrams in both 1996 and 2008, respectively, followed by a calculation of the peak wavenumber of the supergranule feature, the typical supergranule size can be estimated (Section~\ref{ss:SgSize}). Comparing the diameter values listed in Table~\ref{Parameters}, on average supergranules were smaller in 2008 than in 1996, despite the size range itself being approximately equal.

As the horizontal velocities are consistently at least ten times the values of the radial components,
they provide most of the power that is seen within the convection spectrum and are thus the best parameter
with which to make inter-year comparisons. However, the high-wavenumber power discrepancy between the two datasets make any direct velocity comparisons difficult, due to different levels of instrument defocus during the respective observation times.

To equalize the data, the Dopplergrams from each year were spectrally filtered with a log-normal distribution. This maintained the power within the supergranule feature while suppressing the power at higher wavenumbers that is influenced by defocusing. The process described in Section~\ref{ss:SgVelComp}  was then repeated for both filtered datasets. As the same process was applied to both datasets, providing suppression of the problematic high wavenumber region, it is felt that this method provides a reliable inter-year comparison. The horizontal velocity values listed in Table~\ref{Parameters} suggest that the horizontal flows are marginally stronger during 2008 than during 1996.

Strips of data near the solar equator were taken from a base Dopplergram and cross-correlated with strips in subsequent Dopplergrams within each day contained within each dataset. The natural logarithm of the correlation values at each time-lag between the two correlated strips was calculated and averaged over all days.  This provided a dataset that could be linearly fitted, the gradient of which provided the decay coefficient from which the 1/e decay time of the supergranule pattern could be calculated. Assuming the 1/e decay times provide a decent estimate of the supergranule lifetimes, as listed in Table~\ref{Parameters}, supergranule lifetimes are no different between 2008 and 1996.

To conclude, while supergranule lifetimes are the same between 2008 and 1996, their sizes tend to be larger during 1996, while their horizontal flows seem to be stronger during 2008. However, the differences seen are not substantial enough to suggest that the convection zone was behaving any differently during the two epochs to reflect the dramatic differences observed within the sunspot number. This is in contrast to other work such as the analyses of \inlinecite{Howe09}, who saw distinct differences within torsional oscillation maps covering both solar minima.

To better determine whether the observed differences do actually correlate to differences within the global field between the two years requires the results to placed into the larger context of data gathered around solar maximum where the magnetic field strength is quite different to that at minimum. An analysis of data from 2009 would also provide better comparisons between minima as the activity during 1996 and 2009 appears to be of similar phase within the solar cycle. These future analyses will improve our picture of the interaction between the magnetic field and convection. Comparisons may also be made with other research into convection cell characteristics throughout a solar cycle such as the supergranule size measurements made by \inlinecite{Meunier07} and to analyses of cell sizes, their relationship to velocities inside the cell and intensity variations across the cell by \inlinecite{Meunier08}.

Further studies will apply methods similar to those described in this paper to individual Dopplergrams and present a statistical analysis of the time-dependent parameters investigated in this paper. The work will be extended to study characteristics of supergranule-influenced features visible in MDI magnetograms, Ca {\sc II} K images of the solar chromosphere using the instruments such as the Precision Solar Photometric Telescope and H$\alpha$ observations. Supergranule velocity comparisons may also be made by extracting their component values using Time-Distance Helioseismology methods \cite{Duvall93}.

The studies of convection cells at cycle minimum within these many regimes of observation may also assist in quantifying hemispheric asymmetries that may exist. A number of similar studies have been made (for example, \opencite{Osherovich99}) and it would be of interest to observe and compare any hemispheric differences that may exist within each of the characteristics studied during the periods of solar minima covered in this paper.

With the launch of the Helioseismic Magnetic Imager (HMI) instrument (J. Schou, private communication) aboard the Solar Dynamics Observatory (SDO) satellite (W.D. Pesnell, private communication) in February 2010, the spatial range of the studies provided here may be extended by the enhanced spatial resolution of HMI. Observations made via the high-resolution mode of MDI provide convection spectra that clearly show a feature due to granulation \cite{Hathaway00}. SDO/HMI offers the opportunity to study granulation at the same resolution as supergranulation and make comparisons of convection cell characteristics throughout this extended convection spectrum.

%

%

%

%
 \begin{acks}
This research was supported by an appointment to the NASA Postdoctoral Program at NASA Goddard
Space Flight Center, administered by Oak Ridge Associated Universities through a contract with NASA via
the Solar Dynamics Observatory. SOHO is a project of international cooperation between ESA and NASA. The authors
extend their gratitude to John Beck of Stanford University for producing the de-rotated Dopplergram datasets. We also thank the
referee for the constructive comments.
 \end{acks}

%
%
%

\end{article}
\end{document}